# TECHNOLOGICAL STRATEGY OF USING GLOBAL POSITIONING SYSTEM: AN ANALYSIS

Dr.S.S.Riaz Ahamed.

Principal, Sathak Institute of Technology, Ramanathapuram,TamilNadu, India-623501.

ssriaz@ieee.org

## ABSTRACT

The Global Positioning System (GPS) is a U.S. space-based radionavigation system that provides reliable positioning, navigation, and timing services to civilian users on a continuous worldwide basis -- freely available to all. GPS provides specially coded satellite signals that can be processed in a GPS receiver, enabling the receiver to compute position, velocity and time. Basically GPS works by using four GPS satellite signals to compute positions in three dimensions (and the time offset) in the receiver clock. GPS provides accurate location and time information for an unlimited number of people in all weather, day and night, anywhere in the world. Anyone who needs to keep track of where he or she is, to find his or her way to a specified location, or know what direction and how fast he or she is going can utilize the benefits of the global positioning system. Everyday activities such as banking, mobile phone operations, and even the control of power grids, are facilitated by the accurate timing provided by GPS.

## 1.INTRODUCTION

The Global Positioning System (GPS) is a network of 24 Navstar satellites orbiting Earth at 11,000 miles. Originally established by the U.S. Department of Defence (DOD) at a cost of about US$13 billion, access to GPS is free to all users, including those in other countries. The system's positioning and timing data are used for a variety of applications, including air, land and sea navigation, vehicle and vessel tracking, surveying and mapping, and asset and natural resource management. With military accuracy restrictions partially lifted in March 1996 and fully lifted in May 2000, GPS can now pinpoint the location of objects as small as a penny anywhere on the earth's surface[1]-[7].

GPS provides specially coded satellite signals that can be processed in a GPS receiver, enabling the receiver to compute position, velocity and time. Basically GPS works by using four GPS satellite signals to compute positions in three dimensions (and the time offset) in the receiver clock. So by very accurately measuring our distance from these satellites a user can triangulate their position anywhere on earth.

Each GPS satellite has an atomic clock, and continually transmits messages containing the current time at the start of the message, parameters to calculate the location of the satellite (the ephemeris), and the general system health (the almanac). The signals travel at the speed of light through outer space, and slightly slower through the atmosphere. The receiver uses the arrival time to compute the distance to each satellite, from which it determines the position of the receiver using geometry and trigonometry.

Although four satellites are required for normal operation, fewer may be needed in some special cases. If one variable is already known (for example, a sea-going ship knows its altitude is 0), a receiver can determine its position using only three satellites. Also, in practice, receivers use additional clues (doppler shift of satellite signals, last known position, dead reckoning, inertial navigation, and so on) to give degraded answers when fewer than four satellites are visible.

A GPS signal contains three different bits of information — a pseudorandom code, ephemeris data and almanac data. The pseudorandom code is simply an I.D. code that identifies which satellite is transmitting information. You can view this number on your Garmin GPS unit's satellite page, as it identifies which satellites it's receiving. Ephemeris data tells the GPS receiver where each GPS satellite should be at any time throughout the day. Each satellite transmits ephemeris data showing the orbital information for that satellite and for every other satellite in the system. Almanac data, which is constantly transmitted by each satellite, contains important information about the status of the satellite (healthy or unhealthy), current date and time. This part of the signal is essential for determining a position [3][6][11][16].

## 2. GPS SYSTEM SEGMENTS

The Global Positioning System is comprised of three segments: satellite constellation, ground control/monitoring network and user receiving equipment. Formal GPS Joint Program Office (JPO) programmatic terms for these components are space, operational control and user equipment segments, respectively.

1. The satellite constellation contains the satellites in orbit that provide the ranging signals and data messages to the user equipment.
2. The operational control segment (OCS) tracks and maintains the satellites in space. The OCS monitors satellite health and signal integrity and maintains the orbital configuration of the satellites. Furthermore, the OCS updates the satellite clock corrections and ephemerides as well as numerous other parameters essential to determining user position, velocity, and time (PVT).
3. Lastly, the user receiver equipment performs the navigation, timing or other related functions (e.g. surveying).

### 2.1 Gps Satellite Constellation

The satellite constellation consists of the nominal 24-satellite constellation (the first was launched in 1978 and the 24$^{th}$ in 1994). They transmit signals (at 1575.42 MHz) that can be detected by receivers on the ground. The satellites are positioned in six Earth-centred orbital planes with four satellites in each plane. This means that signals from six of them can be received 100 percent of the time at any point on earth. The nominal orbital period of a GPS satellite is one half of a sidereal day or 11 hr 58 min. The orbits are nearly circular and equally spaced about the equator at a 60° degree separation with an inclination relative to the equator of nominally 55° degrees. The orbital radius is approximately 26,600 km (i.e., distance from satellite to centre of mass of the earth).

GPS satellites transmit two low power radio signals, designated L1 and L2. Civilian GPS uses the L1 frequency of 1575.42 MHz in the UHF band.A GPS signal contains three different bits of information — a pseudo-random code, ephemeris data and almanac data. The pseudo-random code is simply an I.D. code that identifies which satellite is transmitting information [19][21][29].

Several different notations are used to refer to the satellites in their orbits. One particular notation assigns a letter to each orbital plane (i.e., A, B, C, D, E, and F) with each satellite within a plane assigned a number from 1 to 4. Thus, a satellite referenced as B3 refers to satellite number 3 in orbital plane B. A second notation used is a NAVSTAR satellite number assigned by the U.S. Air Force. This notation is in the form of a space vehicle number (SVN) 11 to refer to NAVSTAR satellite 11.

### 2.2 Operational Control Segment (OCS)

The OCS has responsibility for maintaining the satellites and their proper functioning. This includes maintaining the satellites in their proper orbital positions (called station keeping) and monitoring satellite subsystem health and status. The OCS also monitors the satellite solar arrays, battery power levels, and propellant levels used for manoeuvres and activates spare satellites. The overall structure of the operational ground/control segment is as follows: Remote monitor stations constantly track and gather C/A and P(Y) code from the satellites and transmit this data to the Master Control Station, which is located at Falcon Air Force Base, Colorado Springs. There is also the ground uplink antenna facility, which provides the means of commanding and controlling the satellites and uploading the navigation messages and other data. The unmanned ground monitor stations are located in Hawaii, Kwajalein in the Pacific Ocean, Diego Garcia in the Indian Ocean, Ascension Island in the Atlantic and Colorado Springs, Continental United States. Ground antennas are located in these areas also. These locations have been selected to maximize satellite coverage.

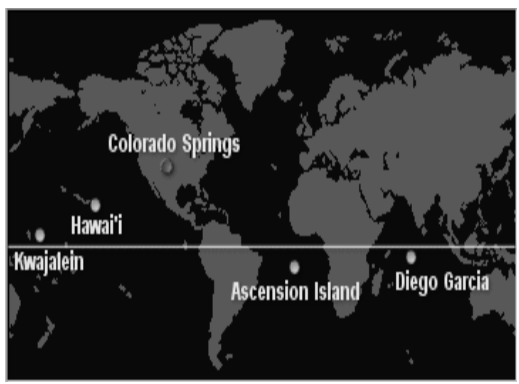

*Location of GPS ground stations*

### 2.3 USER RECEIVING EQUIPMENT

The user receiving equipment, typically referred to as a GPS receiver, processes the L-band signals transmitted from the satellites to determine PVT. There has been a significant evolution in the technology of GPS receiving sets since they were initially manufactured in the mid-70's. Initially, they were large, bulky and heavy analog devices primarily used for military purposes. With today's technology, a GPS receiver of comparable or more capability typically weighs a few pounds or ounces, and occupies a small volume. The smallest of today's are those of a wrist watch size, while the largest is a naval shipboard unit (weighing about 32 kgs). The basic structure of a receiver is the antenna, the receiver and processor, the display and a regulated dc-power supply. These receivers can be mounted in ships, planes and cars, and provide exact position information, regardless of weather conditions[26]-[30][31].

### 3 GPS SYSTEM OPERATION

The basic idea behind GPS is to use satellites in space as reference points for locations on earth. With GPS, signals from the satellites arrive at the exact position of the user and are triangulated. This triangulation is the key behind accurate location determining and is achieved through several steps.

### 3.1 Determining Your Position

Suppose we measure our distance from a satellite and find it to be 11,000 miles (how it is measured is covered later). Knowing that we're 11,000 miles from a particular satellite narrows down all the possible locations we could be in the whole universe to the surface of a sphere that is centered on this satellite and has a radius of 11,000 miles.

Next, say we measure our distance to a second satellite and find out that it's 12,000 miles away. That tells us that we're not only on the first sphere but we're also on a sphere that's 12,000 miles from the second satellite, i.e. somewhere on the circle where these two spheres intersect. If we then make a measurement from a third satellite and find that we're 13,000 miles from that one, that narrows our position down even further, to the two points where the 13,000 mile sphere cuts through the circle that's the intersection of the first two spheres.

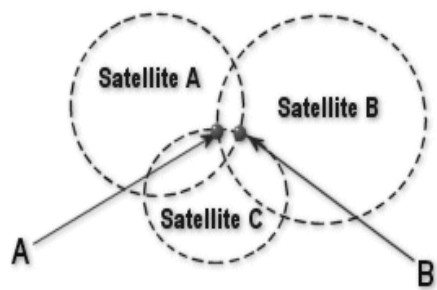

*The two possible locations*

So by ranging from three satellites we can narrow our position to just two points in space. To decide which one is our true location we could make a fourth measurement. But usually one of the two points is a ridiculous answer (either too far from Earth or moving at an impossible velocity) and therefore can be rejected without a measurement.

### 3.2 Measuring Your Distance

How the satellites actually measure the distance is quite different from determining your position and essentially involves using the travel time of a radio message from the satellite to a ground receiver. To make the measurement we assume that both the satellite and our receiver are generating the same pseudo-random code at exactly the same time. This pseudo-random code is a digital code unique to each satellite, designed to be complex enough to ensure that the receiver doesn't accidentally sync up to some other signal. Since each satellite has its own unique Pseudo-Random Code this complexity also guarantees that the receiver won't accidentally pick up another satellite's signal. So all the satellites can use the same frequency without jamming each other. And it makes it more difficult for a hostile force to jam the system, as well as giving the DOD a way to control access to the system.

By comparing how late the satellite's pseudo-random code appears compared to our receiver's code, we determine how long it took to reach us. Multiply that travel time by the speed of light and you obtain the distance between the receiver and the satellite. However this calls for precise timing to determine the interval between the code being generated at the receiver and received from space. On the satellite side, timing is almost perfect due to their atomic clocks installed within each satellite. However as it would be extremely uneconomical for receiver to use atomic clocks a different method must be found.

GPS solves this problem by using an extra satellite measurement for the following reason: If our receiver's clocks were perfect, then all our satellite ranges would intersect at a single point - our position. But with imperfect clocks, a fourth measurement, will not intersect with the first three satellite ranges. So the receiver's computer will then calculate a single correction factor that it can subtract from all its timing measurements that would cause them all to intersect at a single point. That correction brings the receiver's clock back into sync with universal time, ensuring (once the correction is applied to all the rest of the receivers' measurements) precise positioning[12][17].

### 3.3 Error Correction

As would be expected, a variety of different errors can occur within the system, some of which are natural, whilst others are artificial. First of all, a basic assumption, the speed of light, is not constant as this value changes as the satellite signals travel through the atmosphere. As a GPS signal passes through the charged particles of the ionosphere and then through the water vapour of the troposphere it gets slowed down, and this creates the same kind of error as bad clocks. This problem is tackled by attempting to use modelling of the atmospheric conditions of the day, and using dual-frequency measurement, i.e. comparing the relative speeds of two different signals. Another problem is multipath error, this is when the signal may bounce off various local obstructions before it gets to our receiver. Sophisticated signal rejection techniques are used to minimize this problem.

There are also potential problems at the satellites. Minute time differences can occur within the on-board atomic clocks, and sometimes position (*ephemeris*) errors can occur. These other errors can be magnified by a high GDOP "*Geometric Dilution of Precision*" This is where a receiver picks satellites that are close together in the sky, meaning the intersecting circles that define a position will cross at very shallow angles. That increases the grey area or error margin around a position. If the receiver picks satellites that are widely separated the circles intersect at almost right angles and that minimises the error region. Obviously good receivers determine which satellites will give the lowest GDOP.

Finally up to recently there was another , man-made source of errors. The U.S. was very mindful of the fact that terrorists and unfriendly governments could use the accurate positioning provided by GPS and so intentionally degraded GPS's accuracy. This policy is called *Selective Availability* or SA. This involves the DOD introducing some "noise" into the satellite's clock data which, in turn, adds noise (or inaccuracy) into position calculations. The DOD may also has been sending slightly erroneous orbital data to the satellites which they transmit back to receivers on the ground as part of a status message. Together these factors made SA the biggest single source of inaccuracy in the system. Military receivers used a decryption key to remove the SA errors and so they were considerably more accurate.However, effective May 2, 2000 selective availability has been eliminated. The recent terrorist attacks on America have not changed this position. This is due to the fact that civilian uses of GPS have become critical across the world, and because the United States Department of Defence now has the technology to localise the control system to deny GPS signals to selected areas.

**3.4 Differential Gps**

Using a modified form of GPS called *Differential GPS* (originally initiated by the U.S. Coast Guard to counter the accuracy degradation caused by Selective Availability) can significantly reduce the above errors. Even with SA eliminated, DGPS continues to be a key tool for highly precise navigation on land and sea. DGPS can yield measurements accurate to a couple of meters in moving applications and even better in stationary situations. Differential GPS involves the co-operation of two receivers, one that's stationary and another that's roving around making position measurements.

As each GPS receivers use timing signals from at least four satellites to establish a position then each of those timing signals is going to have some error or delay depending on what sort of problems have occurred it on its journey down to Earth. Since each of the timing signals that go into a position calculation has some error, that calculation is going to be a compounding of those errors.

However if two receivers are fairly close to each other, say within a few hundred kilometres, the signals that reach both of them will have travelled through virtually the same slice of atmosphere, and so will have virtually the same errors.

This means that you could use have one receiver to measure the timing errors and then provide correction information to the other receivers that are roving around. This allows virtually all errors to be eliminated from the system.

The reference station operates by receiving the same GPS signals as the roving receiver but instead of working like a normal GPS receiver it uses its known position to calculate timing, rather than using timing signals to calculate position. Essentially determining what the travel time of the GPS signals should be, and compares it with what they actually are. The difference is an "error correction" factor. The receiver then transmits this error information to the roving receiver so it can use it to correct its measurements.

Since the reference receiver has no way of knowing which of the many available satellites a roving receiver might be using to calculate its position, the reference receiver quickly runs through all the visible satellites and computes each of their errors. Then it encodes this information into a standard format and transmits it to the roving receivers. The roving receivers can then apply the corrections for particular satellites they are using. The United States Coast Guard and other international agencies are establishing reference stations all over the place, especially around busy harbours and waterways.

There are also different kinds of DGPS, for use when users don't need precise positioning immediately. This is termed *Post Processing DGPS*, and is used when the roving receiver just needs to record all of its measured positions and the exact time it made each measurement. Then later, this data can be merged with corrections recorded at a reference receiver for a final clean-up of the data, meaning you don't need the radio link required in real-time systems. Another form of DGPS, called *Inverted DGPS*, which is used to save money when operating a large fleet of users. With an inverted DGPS system the users would be equipped with standard GPS receivers and a transmitter and would transmit their standard GPS positions back to the tracking station (the main office). Then at the tracking station the corrections would be applied to the received positions[5]-[9].

### 3.5 Carrier-Phase Gps

This is a new version of GPS that can eliminate errors even better than other forms. Recall that a GPS receiver determines the travel time of a signal from a satellite by comparing the pseudo random code it's generating, with an identical code in the signal from the satellite. The receiver slides its code later and later in time until it syncs up with the satellite's code. The amount it has to slide the code is equal to the signal's travel time. The problem is that the bits (or cycles) of the pseudo random code are so wide that when the signals sync up there is room for error. Survey receivers are better as they start with the pseudo random code and then move on to measurements based on the carrier frequency for that code. This carrier frequency is much higher so its pulses are much closer together and therefore more accurate. At the speed of light the 1.57 GHz GPS signal has a wavelength of roughly twenty centimetres, so the carrier signal can act as a much more accurate reference than the pseudo random code by itself. And if it can get to within one percent of perfect phase like you expect with code-phase receivers you can (theoretically) obtain 3 or 4 millimetre accuracy.

In essence this method is counting the exact number of carrier cycles between the satellite and the receiver. The problem is that the carrier frequency is hard to count because it's so uniform. Every cycle looks like every other. The pseudo random code on the other hand is intentionally complex to make it easier to know which cycle you're looking at. But Carrier-phase GPS tackles this problem by using code-phase techniques to get close. If the code measurement can be made accurate to say, a meter, then we only have a few wavelengths of carrier to consider as we try to determine which cycle really marks the edge of our timing pulse. Resolving this carrier phase ambiguity for just a few cycles is a much more tractable problem and as the computers inside the receivers increase in processing power and functionality it's becoming possible to make this kind of measurement without all the steps that survey receivers go through[30]-[35].

### 4. ATMOSPHERIC EFFECTS

The GPS signals passing through the atmosphere encounter refraction effects including ray bending and propagation delays. These include the atmospheric effects of the troposphere and ionosphere.

### 4.1 Troposphere

The largest effects of the troposphere can be avoided by prescribing an elevation mask for your receiver, thereby avoiding signals from low elevation satellites. With a 15 degree elevation mask, 4-8 satellites will be simultaneously observable from a location on the Earth at any instant of time. The troposphere is composed of the "hydrostatic (dry)" portion and the "wet" portion accounting for water vapor. The dry portion constitutes 90% of the tropospheric refraction, whereas the wet portion constitutes 10%. However, the models for the dry troposphere are more accurate than the models for the wet troposphere. Therefore, the errors in the wet troposphere have a larger effect on the pseudorange bias than the errors in the dry troposphere.

### 4.2 Ionosphere

Some models try to account for all effects of the ionosphere, but require much effort in modeling the highly time dependent total electron count of the atmosphere. A technique to remove the first order effects of the ionosphere linearly combines the L1 and L2 observables to form a new signal that is free of ionospheric effects. Alternatively, a correction to one of the two signals can be solved for. The first order contribution of the ionosphere to the pseudorange bias is related to the inverse of the frequency squared. Thus for the two pseudoranges:

$$\rho_{L1} = \rho_{TRUE} + c(d_s - d_R) + \frac{a}{f_{L1}^2}$$

$$\rho_{L2} = \rho_{TRUE} + c(d_s - d_R) + \frac{a}{f_{L2}^2}$$

We can form an ionosphere free pseudorange by taking a linear combination to cancel the effects which results in an ionospheric free pseudorange observable of:

$$\rho_{IONO} = \rho_{L1} - \frac{f_{L2}^2}{f_{L1}^2} \rho_{L2}$$

A similar development exists for the carrier phase observable. The expressions for the phase derived pseudoranges with first order ionosphere corrections are:

$$\Phi_{IONO} = \Phi_{L1} - \frac{f_{L2}}{f_{L1}} \Phi_{L2}$$

To actually use this new observable, it is necessary to compute the wavelength of the ionosphere free signal. Substituting values in yields a wavelength of about 48.5 cm for the ionosphere free signal or a frequency of 618.8 Mhz (60.5*f.0). Because the ratio between the L1 and L2 frequencies is not an integer value, the ambiguity term is no longer an integer. Alternatively, the ionospheric correction, a, could have been solved for.

**5.GPS APPLICATIONS**

The Global Positioning System, while originally a military project, is considered a *dual-use* technology, meaning it has significant applications for both the military and the civilian industry.

**5.1 Military**

- Target tracking: Various military weapons systems use GPS to track potential ground and air targets before they are flagged as hostile.These weapon systems pass GPS co-ordinates of targets to precision-guided munitions to allow them to engage the targets accurately.
- Navigation: GPS allows soldiers to find objectives in the dark or in unfamiliar territory, and to coordinate the movement of troops and supplies. The GPS-receivers commanders and soldiers use are respectively called the **Commanders Digital Assistant** and the **Soldier Digital Assistant**.
- Missile and projectile guidance: GPS allows accurate targeting of various military weapons including ICBMs, cruise missiles and precision-guided munitions. Artillery projectiles with embedded GPS receivers able to withstand accelerations of 12,000G have been developed for use in 155 mm howitzers.
- Search and Rescue: Downed pilots can be located faster if they have a GPS receiver.
- The GPS satellites also carry a set of nuclear detonation detectors consisting of an optical sensor (Y-sensor), an X-ray sensor, a dosimeter, and an Electro-Magnetic Pulse (EMP) sensor (W-sensor) which form a major portion of the United States Nuclear Detonation Detection System.

**5.2 Civilian**

Many civilian applications benefit from GPS signals, using one or more of three basic components of the GPS: absolute location, relative movement, and time transfer.

The ability to determine the receiver's absolute location allows GPS receivers to perform as a surveying tool or as an aid to navigation. The capacity to determine relative movement enables a receiver to calculate local velocity and orientation, useful in vessels or observations of the Earth. Being able to synchronize clocks to exacting standards enables time transfer, which is critical in large communication and observation systems. An example is CDMA digital cellular. Each base station has a GPS timing receiver to synchronize its spreading codes with other base stations to facilitate inter-cell hand off and support hybrid GPS/CDMA positioning of mobiles for emergency calls and other applications. Finally, GPS enables researchers to explore the Earth environment including the atmosphere, ionosphere and gravity field. GPS survey equipment has revolutionized tectonics by directly measuring the motion of faults in earthquakes.

To help prevent civilian GPS guidance from being used in an enemy's military or improvised weaponry, the US Government controls the export of civilian receivers. A US-based manufacturer cannot generally export a GPS receiver unless the receiver contains limits restricting it from functioning when it is simultaneously (1) at an altitude above 18 kilometers (60,000 ft) and (2) traveling at over 515 m/s (1,000 knots).[48] These parameters are well above the operating characteristics of the typical cruise missile, but would be characteristic of the reentry vehicle from a ballistic missile[3][11][31].

## 6. CONCLUSIONS

GPS functionality has now started to move into mobile phones. GPS has a variety of applications on land, at sea and in the air. Basically, GPS is usable everywhere except where it's impossible to receive the signal such as inside most buildings, in caves and other subterranean locations, and underwater. The most common airborne applications are for navigation by general aviation and commercial aircraft. At sea, GPS is also typically used for navigation by recreational boaters, commercial fishermen, and professional mariners. Land-based applications are more diverse. The scientific community uses GPS for its precision timing capability and position information. Surveyors use GPS for an increasing portion of their work. GPS offers cost savings by drastically reducing setup time at the survey site and providing incredible accuracy. Basic survey units, costing thousands of dollars, can offer accuracies down to one meter. More expensive systems are available that can provide accuracies to within a centimeter. Recreational uses of GPS are almost as varied as the number of recreational sports available. GPS is popular among hikers, hunters, snowmobilers, mountain bikers, and cross-country skiers, just to name a few.

## 7.REFERENCES

[1]. Hoffman-Wellenhoff, B., H. Lichtenegger, and J. Collins. *Global Positioning*. 3rd Ed. New York: Springer-Verlag Wien, 1994,Pp 110-195.
[2]. Jensen, M.H. *Quality Control for Differential GPS in Offshore Gas and Oil Exploration*. GPS World 3, no8, pp.36-48.
[3]. I. A. Getting, "The Global Positioning System," *IEEE Spectrum*, December 1993, pp. 36-47.
[4]. Y. Hada and K. Takase, "Multiple Mobile Robot Navigation Using the Indoor Global Positioning System (iGPS)," *Proc. IEEE/RSJ Conf. Intelligent Robots and Syst.*, pp. 1005-1010.
[5]. J. Hightower and G. Borriello, "Location Systems for Ubiquitous Computing," *IEEE Computer Magazine*, August 2001, pp. 57-66.
[6]. N. F. Krasner et al., "Position Determination Using Hybrid GPS/Cellphone Ranging," Inst. of Navigation Conf. GPS 2002.
[7]. Parkinson, B.W. (1996), *Global Positioning System: Theory and Applications*, chap. 1: Introduction and Heritage of NAVSTAR, the Global Positioning System. pp. 3-28, American Institute of Aeronautics and Astronautics, Washington, D.C.
[8]. Kaplan, E. *Understanding GPS: Principles and Applications*. Norwood, Conn.: Artech House, 1996,Pp 79-123.
[9]. Bauer, W.D. and M. Schefcik. "Using Differential GPS to improve Crop Yields" GPS World. February 1994,5, no2, pp.38-41.
[10]. Fossum, Donna, David Frelinger,Gerald Frost, Irving Lachow, Scott Pace, and Monica Pinto. *The Global Positioning System*. Santa Monica, Ca.: Rand, 1995.
[11]. National Academy of Public Administration and National Research Council. *Charting the Future: The Global Positioning System*. Washington, D.C.: Sherwood Fletcher Associates, 1995.
[12]. Peterson, Julie. *Understanding Surveillance Technology*. Boca Raton, Fla.:CRC Press, 2001.
[13]. Teunissen, P.J.G. and A. Kleusberg., eds. *GPS for Geodesy*. 2nd Ed. New York: Springer-Verlag, 1998.
[14]. R. Allan, "Onstar System Puts Telematics on the Map," *Electronics Design*, 31 March 2003, pp. 49-56.
[15]. P. Bahl and V. N. Padmanabhan, "RADAR: An In-Building RF-based User Location and Tracking System," *Proc. IEEE Infocom 2000*, pp. 775-784.
[16]. J. Blyer, "Location-Based Services Are Positioned for Growth," *Wireless Systems Design*, September 2003, pp. 16-20.
[17]. J. J. Caffery and G. L. Stuber, "Overview of radiolocation in CDMA cellular systems," *IEEE Communications Magazine*, April 1998, pp. 38-45.

[18]. S Capkun et al, "GPS-free Positioning in Mobile Ad-Hoc Networks," *Proc. IEEE 2001 Hawaii Conf. Syst. Sci.*.
[19]. N. Davies et al., "Using and Determining Location in a Context-Sensitive Tour Guide," *IEEE Computer Magazine*, August 2001, pp. 35-41.
[20]. G. M. Djuknic and R. E. Richton, "Geolocation and Assisted GPS," *IEEE Computer Magazine*, February 2001, pp. 123-125.
[21]. E. Elnahrawy, et al., "The Limits of Localization Using Signal Strength: A Comparative Study," *Proc. IEEE SECON 2004*.
[22]. D. Fox et al., "Markov Localization for Mobile Robots in Dynamic Environments," *J. Artificial Intelligence Research*, Vol. 11 (1999), pp. 391-427.
[23]. H. W. Gellersen, et al., "Multi-Sensor Context-Awareness in Mobile Devices and Smart Artefacts," *Mobile Networks and Applications* (MONET), October 2002.
[24]. A. Ladd et al., "Robotics-Based Location Sensing Using Wireless Ethernet," *Proc. ACM MobiCom 2002*.
[25]. D. Niculescu and B. Nath, "Ad Hoc Positioning System (APS)," *Proc. IEEE Globecom 2001*, pp. 2926-2931.
[26]. N. Patwari et al., "Relative Location in Wireless Networks," *Proc. IEEE Spring VTC 2001*, pp. 1149-1153.
[27]. E. Prigge and J. How, "An Indoor Absolute Positioning System with no Line of Sight Restrictions and Building-Wide Coverage," *Proc. 2000 IEEE Conf. Robotics and Automation*, pp. 1015-1022.
[28]. N. B. Priyantha et al., "The Cricket Location-Support System," *Proc. ACM MobiCom 2002*.
[29]. T. S. Rappaport, et al., "Position Location Using Wireless Communications on Highways of the Future," *IEEE Communications Magazine*, October 1996, pp. 33-41.
[30]. C. Savarese et al., "Locationing in Distributed Ad-Hoc Wireless Sensor Networks," *Proc. IEEE ICASSP 2001*, pp. 2037-2040.
[31]. A. Savvides et al., "The Bits and Flops of the N-hop Multilateration Primitive for Node Localization Problems," *Proc. ACM WSNA 2002*, pp. 112-121.
[32]. Y. Shang et al., "Localization from Mere Connectivity," *Proc. MobiHoc 2003*, pp. 201-212, June 2003.
[33]. J. Werb and C. Lanzi, "Designing a Positioning System for Finding Things and People Indoors," *IEEE Spectrum*, September 1998, pp. 71-78.
[34]. K. Whitehouse and D. Culler, "Calibration as Parameter Estimation in Sensor Networks," *Proc. ACM WSNA 2002*, pp. 59-67.
[35]. Y. Zhao, "Standardization of Mobile Phone Positioning for 3G Systems," *IEEE Communications Magazine*, July 2002, pp. 108-116.